\documentstyle[twoside,fleqn,npb,epsfig]{article}
\newcommand{\AmS}{{\protect\the\textfont2
  A\kern-.1667em\lower.5ex\hbox{M}\kern-.125emS}}

\title{Two-loop matching conditions for $\overline{\rm MS}$ parton densities}

\author{J. Smith\address{C.N. Yang Institute for Theoretical Physics,
        SUNY at Stony Brook , \\
        Stony Brook, NY 11794-3840, USA}%
        \thanks{Work supported in part by NSF PHY-9722101}}

\begin{document}

\begin{abstract}
We discuss how the the operator product expansion 
(OPE) can be used to derive asymptotic expressions for
certain integrals. This yields operator matrix elements (OME's)
which determine the matching conditions for 
$\overline{\rm MS}$ parton densities across 
heavy flavour thresholds. Then we construct four and five-flavour 
densities from a three-flavour set via the evolution of the AP 
equation using LO and NLO splitting functions. 

\end{abstract}

\maketitle

It is well known that $\alpha_s(\mu^2,n_f,\Lambda(n_f))$
in pQCD requires matching conditions as the scale $\mu$
crosses flavour matching points.
At these points the number of light-flavours $n_f$ changes by unity
so the QCD scale parameter $\Lambda(n_f)$ in the solution of the
differential equation for the $\beta$ function 
is redefined to make the running coupling continuous. 
When heavy quarks are included another scale $m$, the mass of the heavy quark,
enters and the matching conditions are more complicated.
The precise relations which need to be satisfied are given in 
\cite{2loop}, \cite{2loop2}.
In order lowest order pQCD one can choose to make the $\alpha_s$
continuous across heavy flavour thresholds at $\mu=m$.
However this does not hold in higher order pQCD 
as the matching conditions in the $\overline{\rm MS}$
scheme then contain non-logarithmic terms.  Hence there is a 
discontinuity in $\alpha_s$ at $\mu=m$.
 
Recently the analogous problem of deriving the two-loop matching conditions
on parton densities as the mass factorization scale crosses the
heavy flavour thresholds has been solved in \cite{bmsn1}.
The way this was done is as follows.
We examined the large $Q^2$ limit of the heavy quark coefficient functions
which appear in NLO perturbation expressions for heavy quark 
extrinsic pair production in deep inelastic scattering. 
These quantities are functions
of the virtuality of the photon probe $\sqrt{Q^2}$, the mass of the heavy quark
$m$, the renormalization scale $\mu$, which is chosen equal to the mass
factorization scale, and the partonic Bjorken scaling variable $z$.
The number of heavy $D^*$ mesons produced in deep inelastic scattering 
can be derived by convoluting these heavy ($c -  \bar c$) quark coefficient 
functions with appropriate combinations of three-flavour light
parton densities (u,d,s and g) and with heavy ($c - \bar c$)  quark 
fragmentation functions \cite{lrsn}. Note that the heavy 
$c - \bar c$ pair only appears in the final state. 
Unfortunately we do not have analytic expressions for all
these heavy quark coefficient functions. Some only
exist as two-dimensional integrals over very complicated expressions. 
However there are convenient tables for all of them in \cite{rsn}. 

In the limit $Q^2\gg m^2$ the complicated integrals in the
heavy quark coefficient functions reduce
to terms with powers of $\ln(Q^2/\mu^2)$ and $\ln(\mu^2/m^2)$ 
multiplied by functions of the variable $z$. 
These results can be reexpressed as convolutions of
light-mass coefficient functions ${\cal C}(z,Q^2/\mu^2)$ which contain
the terms with powers in $\ln(Q^2/\mu^2)$ and 
OME's $A(z,\mu^2/m^2)$
which contain the powers in $\ln(\mu^2/m^2)$ . The way we 
evaluated these OME's is described in \cite{willy} so we only 
give an outline here.  We wrote the heavy quark coefficient functions 
in terms of dispersion integrals for off-shell forward Compton 
scattering as is normally done for the OPE in deep inelastic scattering.
We then changed
variables to write the dispersion integral in terms of a variable
$z'$ which is between zero and unity. Next we expanded 
the denominator in a Taylor series
in $z'$. To take the limit $Q^2 \gg m^2$ of the dispersion integral 
we add and subtract the same dispersion integral where we take the limit
$Q^2 \gg m^2$ in the integrand. This integrand contains the 
OPE of the standard heavy quark (Q) nonsinglet and 
singlet operators in pQCD taken between states
with momentum $k$, namely
\begin{eqnarray}
<Q(k)| O_{Q,\mu_1,\mu_2,.... \mu_n}\,(0)|Q(k)> \,.
\end{eqnarray}
The heavy quark operator 
\begin{eqnarray}
O_{Q,\mu_1,\mu_2,.... \mu_n}(x) = \bar \psi(x)\gamma_{\mu_1}
D_{\mu_2}......D_{\mu_n} \psi(x) \,,
\end{eqnarray}
is a gauge invariant operator containing 
the heavy quark field $\psi(x)$ and the covariant
derivative $D_{\mu} = \partial_{\mu} + i g A_{\mu}$.
It can be shown that the original integral minus the integral 
involving the OPE does not contain any mass
singularities as $m \rightarrow 0$ so it cannot depend on the heavy quark
mass $m$ and therefore only contains terms with powers of $\ln(Q^2/\mu^2)$. 
Hence the integrals which contain the evaluation of the 
OME's in the OPE yield all the terms containing powers of
$\ln(\mu^2/m^2)$. This means that analytic expressions for the
two-loop OME's with one heavy quark loop 
and light-quark or gluon incoming and outgoing states
contain the information we require to extract the 
$A(z,\mu^2/m^2)$.
Of course the actual evaluation of the five OME's which exist in order
$\alpha_s^2$ 
requires the introduction of infrared and ultraviolet regulators,
the use of gauge invariant operators, contractions with light-like
four vectors to make the projections and $\overline{\rm MS}$ 
renormalization. The results of this analysis are encapsulated in 
expressions like 
\begin{eqnarray}
&&\tilde A^{{\rm PS},(2)}_{Qq}(z, \mu^2/m^2) = A_1(z) \ln^2(\mu^2/m^2) 
\nonumber\\
&&+ A_2(z) \ln(\mu^2/m^2) + A_3(z)\,,
\end{eqnarray}
where 
\begin{eqnarray}
&&A_1(z) = C_F T_f[-8(1+z)\ln{z} 
\nonumber\\
&&
-16/(3z) -4+4z+16z^2/3]\,, 
\end{eqnarray}
\begin{eqnarray}
&& A_2(z) =C_F T_f [-8(1+z)\ln^2{z} 
\nonumber\\
&&
+(8+40z+64z^2/3)\ln{z}
\nonumber\\
&&
+160/(9z) -16+48z-448z^2/9]\,,
\end{eqnarray}
\begin{eqnarray}
&& A_3(z) =C_F T_f\{(1+z)[32{\rm S}_{1,2}(1-z) 
\nonumber\\
&&
+16\ln{z}{\rm Li}_2(1-z)
\nonumber\\
&&
-16\zeta(2)\ln{z} -4/3\ln^3{z}]
\nonumber\\
&&
+(32/(3z) +8-8z-32z^2/3){\rm Li}_2(1-z)
\nonumber\\
&&
+(-32/(3z)-8+8z+32z^2/3)\zeta(2)
\nonumber\\
&&
+(2+10z+16z^2/3)\ln^2{z}
\nonumber\\
&&
-(56/3+88z/3+448z^2/9)\ln{z} -448/(27z)
\nonumber\\
&&
 -4/3 -124z/3+1600z^2/27\} \,.
\end{eqnarray}
The tilde indicates that an overall factor of $n_f$ has been extracted
from the function and the (2) in the superscript
means this is the second order term in an expansion in $a_s=\alpha_s /(4\pi)$.
The five functions 
$\tilde A^{{\rm PS},(2)}_{Qq}(z, \mu^2/m^2)$,
where PS denotes pure singlet under the flavour group
(i.e., no non-singlet projection exists), 
$\tilde A^{{\rm S},(2)}_{Qg}(z, \mu^2/m^2)$,
where S denotes singlet under the flavour group,
$A^{{\rm S},(2)}_{gg,Q}(z, \mu^2/m^2)$,
$A^{{\rm S},(2)}_{gq,Q}(z, \mu^2/m^2)$,
and $A^{{\rm NS},(2)}_{qq,Q}(z, \mu^2/m^2)$ where NS denotes
non-singlet under the flavour group,
which exist 
in order $\alpha_s^2$ pQCD are given in \cite{bmsn1}. Alternative discussions
of their derivation and use are given in \cite{bmsn2}.
Note that they contain nonlogarithmic terms such as $A_3(z)$
in Eq.(6) in order $\alpha_s^2$ so
there is no scale $\mu$ where we can make them all vanish.
Since we know the four-flavour light mass coefficient functions 
${\cal C}(z,Q^2/\mu^2)$ in order $\alpha_s^2$ \cite{zn1}
we can analytically evaluate the convolutions with the appropriate
$A$'s to obtain asymptotic expressions for the heavy quark coefficient 
functions. They were given in \cite{bmsmn}.
As far as this workshop is concerned we would like
to point out that this "inverse mass factorization method"
is an elegant use of the OPE to obtain asymptotic
expansions of integrals. 

\begin{figure}[ht]
   \epsfig{file=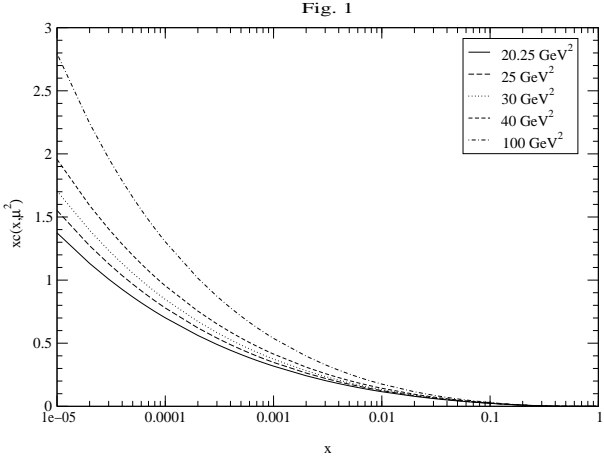,width=6cm,angle=270 }
\vspace{-0.6cm}
\caption{$xc_{\rm NNLO}(5,x,\mu^2)$ in the range
$10^{-5} < x < 1$ for $\mu^2 =$ 20.25, 25, 30, 40 and 100 in units
of $({\rm GeV/c}^2)^2$.}
\vspace{-0.6cm}
\end{figure}                                                              

Normally parton densities are fitted to specific functions of $x$ at a 
scale $\mu$ and the AP equations then govern the evolution of these
densities to other scales. Suppose one begins with a three-flavour set 
containing densities for u,d,s quarks and the gluon g. Then
the above results allow one to define four-flavour parton densities 
at scales $\mu \ge m_c$ from the input set of three-flavour densities 
in fixed-order perturbation theory (FOPT). 
Let the $\otimes$ symbol denote the convolution integral
$f\otimes g=\int f(x/y)g(y)dy/y$, where $x \le y \le 1$,
then we define the charm density  
\begin{eqnarray}
&&f_{c + {\bar c}}(n_f+1,\mu^2)=
\nonumber\\
&& a_s(n_f,\mu^2) \tilde A_{Qg}^{\rm S}\Big(\frac{\mu^2}{m_c^2}\Big) \otimes
f_g^{\rm S}(n_f, \mu^2) 
\nonumber\\
&& + a_s^2 (n_f,\mu^2)\Big [\tilde A_{Qq}^{\rm PS}
\Big(\frac{\mu^2}{m_c^2}\Big)\otimes f_q^{\rm S}(n_f, \mu^2)
\nonumber\\[2ex]  
&& + \tilde A_{Qg}^{\rm S}\Big(\frac{\mu^2}{m_c^2}\Big) 
\otimes f_g^{\rm S}(n_f, \mu^2)\Big ] \,,
\end{eqnarray}
the singlet gluon density
\begin{eqnarray}
&&f_g^{\rm S}(n_f+1, \mu^2) = f_g^{\rm S}(n_f, \mu^2)
\nonumber\\
&& +a_s(n_f,\mu^2)A_{gg,Q}^{\rm S}(\frac{\mu^2}{m_c^2}) \otimes 
f_g^{\rm S}(n_f, \mu^2)
\nonumber\\
&& + a_s^2(n_f,\mu^2) \Big [ A_{gq,Q}^{\rm S}(\frac{\mu^2}{m_c^2})
\otimes f_q^{\rm S}(n_f, \mu^2) \,, 
\nonumber\\
&&+ A_{gg,Q}^{\rm S}(\frac{\mu^2}{m_c^2})
\otimes f_g^{\rm S}(n_f, \mu^2) \Big ]\,,
\end{eqnarray}
and the light mass quark densities
\begin{eqnarray}
&&f_{k+\bar k}^{\rm}(n_f+1,\mu^2) = f_{k+\bar k}(n_f,\mu^2) 
\nonumber\\
&& + a_s^2(n_f,\mu^2) A_{qq,Q}^{\rm NS}\Big(\frac{\mu^2}{m_c^2}\Big)
\otimes f_{k+\bar k}(n_f, \mu^2)\,,
\end{eqnarray}
for $n_f = 3$ and $m_c^2 \le \mu^2 <  m_b^2$. 
Note that we have suppressed the $x$ dependence to make 
the notation more compact. These expressions were used in \cite{bmsn2} 
to construct a variable flavour number scheme (VFNS)
for the heavy quark contributions to the deep inelastic structure functions.
\begin{figure}[ht]
   \epsfig{file=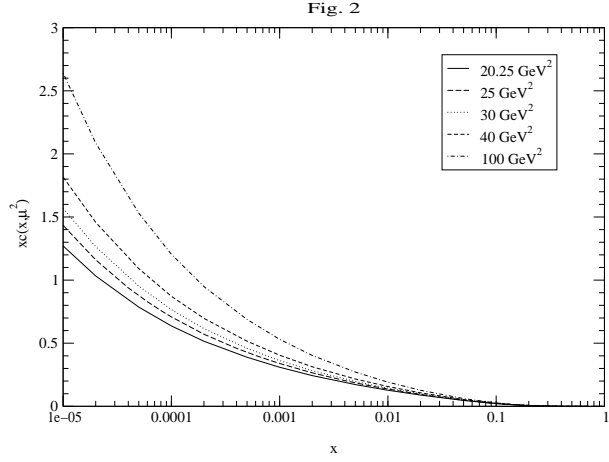,width=6cm,angle=270 }
\vspace{-0.6cm}
\caption{Same as Fig.1 for the NLO results from MRST98 set 1.}
\vspace{-0.6cm}
\end{figure}                                                              
\begin{figure}[ht]
   \epsfig{file=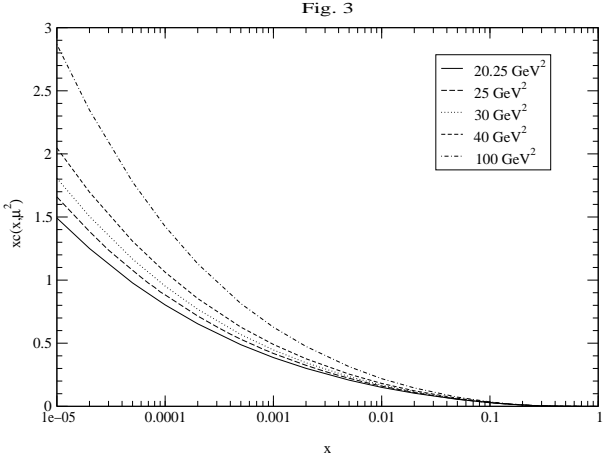,width=6cm,angle=270 }
\vspace{-0.6cm}
\caption{Same as Fig.1 for the NLO results from CTEQ5HQ.} 
\vspace{-0.6cm}
\end{figure}                                                              

Note however that the above procedure does not resum the potentially
large terms in $\ln(\mu^2/m_c^2)$ which are explicitly left in 
the parton densities. To do this we need to evolve   
the above densities via the AP equation rather than using FOPT.
This is new work in \cite{chsm} using three-flavour densities at
small scales from \cite{grv98}. The latter LO and NLO densities are started
at very small scales $\mu_0$ below the mass of the charmed quark.
Hence three flavor evolution proceeds from the initial $\mu_0^2$ to the
scale $\mu^2=m_c^2=1.96$ $({\rm GeV/c}^2)^2$.
In this region $\alpha_s$ is large so we had to be very
careful to get numerically accurate solutions of the evolution equation. 
Fortunately there are standard
inputs and tables in \cite{brnv} with which we could compare the 
parton densities from our evolution code.
We chose the matching scale $\mu$
at the mass of the charm quark $m_c$ so that all the $\ln(\mu^2/m_c^2)$
terms in the OME's vanish at this point leaving only the
nonlogarithmic pieces in the order $\alpha_s^2$ OME's to contribute to
the right-hand-sides of Eqs. (7), (8) and (9). 
Note that the LO and NLO charm densities vanish at the scale $\mu=m_c$
since 
\begin{eqnarray}
\tilde A^{{\rm S},(1)}_{Qg}(z,\mu^2/m^2)
= 4 T_f (z^2 + (1-z)^2)\ln(\mu^2/m^2) \,,
\nonumber\\
\end{eqnarray}
does not have a non-logarithmic term. 
The NNLO charm density starts off with
a finite $x$-dependent shape in order $a_s^2$ determined by 
\begin{eqnarray}
&&f_{c + {\bar c}}(n_f+1,m_c^2)=
\nonumber\\[2ex]
&&a_s^2 (n_f,m_c^2)\Big [\tilde A_{Qq}^{\rm PS}
(1)\otimes f_q^{\rm S}(n_f, m_c^2)
\nonumber\\[2ex]
&& + \tilde A_{Qg}^{\rm S}(1)
\otimes f_g^{\rm S}(n_f, m_c^2)\Big ] \,,
\end{eqnarray}
with $n_f=3$. Hence the OME's provide the boundary condition
for the evolution of the (massless) charm density.
Also note that we ordered the terms on the right-hand-side of Eq.(11) 
in powers of $\alpha_s$ so that the result contains
a product of NLO OME's and LO parton densities,
although this is not evident here. The result is then strictly
order $a_s^2$ and should be multiplied by order $a_s^0$ coefficient
functions when forming the zero-mass variable flavour number
scheme (ZM-VFNS) charm density contribution to the deep 
inelastic structure functions.
\begin{figure}[ht]
   \epsfig{file=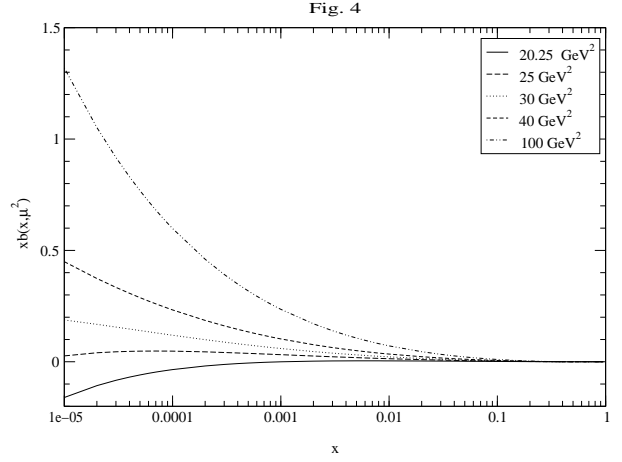,width=6cm,angle=270 }
\vspace{-0.6cm}
\caption{$xb_{\rm NNLO}(5,x,\mu^2)$ in the range
$10^{-5} < x < 1$ for $\mu^2 =$ 20.25, 25, 30, 40 and 100 in units
of $({\rm GeV/c}^2)^2$.}
\vspace{-0.6cm}
\end{figure}                                                              

The four-flavour gluon density is also generated at the
matching point in the same way. At $\mu = m_c$ we define
\begin{eqnarray}
&&f_g^{\rm S}(n_f+1, m_c^2) = f_g^{\rm S}(n_f, m_c^2)
\nonumber\\
&&  + a_s^2(n_f,m_c^2) \Big [ A_{gq,Q}^{\rm S}(1)
\otimes f_q^{\rm S}(n_f, m_c^2) \,, 
\nonumber\\
&&+ A_{gg,Q}^{\rm S}(1)
\otimes f_g^{\rm S}(n_f, m_c^2) \Big ]\,.
\end{eqnarray}
The four-flavor light quark (u,d,s) densities 
are generated using
\begin{eqnarray}
&&f_{k+\bar k}^{\rm}(n_f+1,m_c^2) = f_{k+\bar k}(n_f,m_c^2) 
\nonumber\\
&& + a_s^2(n_f,m_c^2) A_{qq,Q}^{\rm NS}(1)
\otimes f_{k+\bar k}(n_f, m_c^2)\,.
\end{eqnarray}
The {\it total} four-flavor singlet quark density  
follows from the sum of Eqs.(11) and (13).

Next the resulting four-flavor densities are evolved from their
boundary values using the four-flavor evolution kernels in the AP equations
in either LO or NLO up to the scale
$\mu^2=20.25$ $({\rm GeV/c}^2)^2$.
The bottom quark density is then generated at this point using
\begin{eqnarray}
&&f_{b + {\bar b}}(n_f+1,m_b^2)=
\nonumber\\
&&a_s^2 (n_f,m_b^2)\Big [\tilde A_{Qq}^{\rm PS}
(1)\otimes f_q^{\rm S}(n_f, m_b^2)
\nonumber\\
&&+ \tilde A_{Qg}^{(\rm S)}(1)
\otimes f_g^{\rm S}(n_f, m_b^2)\Big ] \,,
\end{eqnarray}
and the five-flavour gluon and light quark densities (which now include charm)
are generated using Eqs. (12) and (13)
with $n_f =4$ and replacing $m_c^2 $ by $m_b^2$.
Therefore only the nonlogarithmic terms in the order $a_s^2$ OME's
contribute to the matching conditions on the bottom quark density.
Then all the densities are evolved up to higher $\mu^2$ as a five-flavor set
with either LO or NLO splitting functions. 

\begin{figure}[ht]
   \epsfig{file=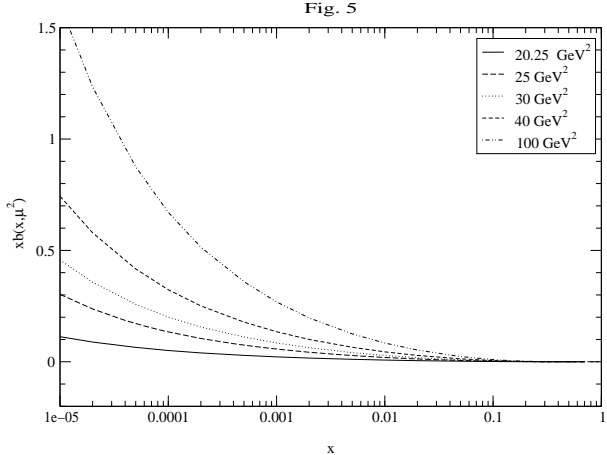,width=6cm,angle=270 }
\vspace{-0.6cm}
\caption{Same as Fig.4 for the NLO results from MRST98 set 1.}
\vspace{-0.6cm}
\end{figure}                                                              

\begin{figure}[ht]
   \epsfig{file=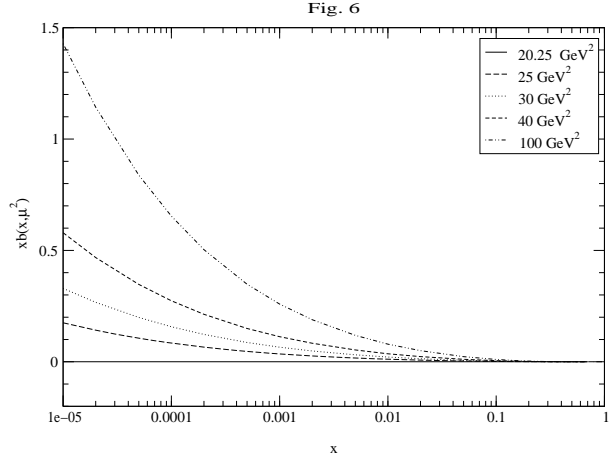,width=6cm,angle=270 }
\vspace{-0.6cm}
\caption{Same as Fig.4 for the NLO results from CTEQ5HQ.}
\vspace{-0.6cm}
\end{figure}                                                              

The above formulae and their evolution with LO and NLO
splitting functions have been implemented in a C++ computer code \cite{chsm}
to yield the CS parton density set. 
They were used in the construction of two VFNS
for the charm quark contribution to the deep inelastic structure functions
in \cite{csn}. Note that approximate expressions for the three loop
splitting functions are now available in \cite{nevo}. When NNLO
parton densities are available from fits to experimental data
we can incorporate them into our computer program.

As an illustration we would like to compare the charm and bottom 
quark densities in the CS \cite{chsm}, MRST98 \cite{mrst98}
and CTEQ5 \cite{cteq5} sets.
The latter two sets work with order $\alpha_s$ matching conditions
so the parton densities are continuous across heavy flavour thresholds.
The MRST98 sets use a procedure proposed in \cite{thro}, while
the CTEQ5 sets use the different ACOT procedure in \cite{acot}.
Here we show the five-flavor densities. 
In the CS set they start at $\mu^2 = m_b^2 = 20.25$ 
${\rm GeV}^2$. At this scale the charm densities in the
CS, MRST98 (set 1) and CTEQ5HQ sets are shown in Figs.1,2,3
respectively. Since the CS charm density starts off negative for small $x$ at
$\mu^2 = m_c^2 = 1.96$ ${\rm GeV}^2$ (see the plots in \cite{chsm})
it is smaller than the corresponding
CTEQ5HQ density. At larger $\mu^2$ all the CS curves in Fig.1 are below 
those for CTEQ5HQ in Fig.3 although the differences are small.
In general the CS c-quark densities
are more equal to those in the MRST98 (set 1) in Fig.2.

At the matching point $\mu^2 = 20.25$ ${\rm GeV}^2$
the b-quark density also starts off negative at small $x$ 
as can be seen in Fig.4, which is a consequence of the explicit
form of the OME's in \cite{bmsn1}. At $O(\alpha_s^2)$ the 
nonlogarithmic terms do not vanish at the matching point
and yield a finite function in $x$, which is the boundary value
for the evolution of the b-quark density.
This negative start slows down the evolution of the b-quark density
at small $x$ as the scale $\mu^2$ increases. Hence the CS densities 
at small $x$ in Fig.4
are smaller than the MRST98 (set 1) densities in Fig.5 and the CTEQ5HQ 
densities in Fig.6 at the same values of $\mu^2$.
The differences between the sets are still small, of the order of five percent
at small $x$ and large $\mu^2$. This will lead to differences
in cross sections for processes
involving incoming b-quarks at the Tevatron.

We suspect that the differences between these results 
for the c and b-quark densities are primarily due to the
different gluon densities in the three sets rather to than the effects
of the different boundary conditions.
This could be checked theoretically if both LO and
NLO three-flavor sets were provided by MRST and CTEQ
at small scales. 
We note that CS uses the GRV98 LO and NLO gluon densities, 
which are rather steep in $x$ and generally
larger than the latter sets at the same values of $\mu^2$.
Since the discontinuous boundary conditions
suppress the charm and bottom densities at small $x$, they enhance 
the gluon densities in this same region (in order that the
momentum sum rules are satisfied).
Hence the GRV98 three flavour gluon densities and the 
CS four and five flavor gluon densities are generally 
larger than those in MRST98 (set 1) and CTEQ5HQ. Unfortunately 
experimental data are not yet precise enough to decide which set is
the best one. 

{\bf Acknowledgements.}  

I am grateful to B. Harris, E. Laenen
and W.L van Neerven for helpful comments on this report.

\end{document}